\documentstyle[prd,aps,preprint,tighten]{revtex}

\begin{document}

\draft
\preprint{\begin{tabular}{r}
{\bf hep-ph/9909514} 
\end{tabular}}

\title{Bi-maximal Neutrino Mixing Pattern Reexamined}
\author{Zhi-zhong Xing}
\address{Sektion Physik, Universit$\it\ddot{a}$t M$\it\ddot{u}$nchen,
Theresienstrasse 37A, 80333 M$\it\ddot{u}$nchen, Germany \\
{\it Electronic address: xing@theorie.physik.uni-muenchen.de} }
\maketitle

\begin{abstract}
I propose a simple extension of the bi-maximal neutrino mixing
pattern, allowing slight coupling between solar and
atmospheric neutrino oscillations as well as large $CP$ violation. 
The new scenario is compatible with either the vacuum oscillation
solution or the large-angle MSW solution to the solar neutrino problem.
\end{abstract}

\pacs{PACS number(s): 14.60.Pq, 13.10.+q, 25.30.Pt}

\newpage

To interpret current experimental data on 
atmospheric and solar neutrino oscillations, a lot of interest 
has been paid to the ``bi-maximal'' neutrino mixing pattern 
(up to a trivial sign rearrangement) \cite{Barger98}:
\begin{equation}
U \; =\; \left ( \matrix{
\frac{1}{\sqrt{2}}	& \frac{1}{\sqrt{2}}	& 0 \cr\cr
-\frac{1}{2}	& \frac{1}{2}	& \frac{1}{\sqrt{2}} \cr\cr
\frac{1}{2}	& -\frac{1}{2}	& \frac{1}{\sqrt{2}} \cr} \right ) \;\; .
\end{equation}
Note that $U$ links the neutrino mass eigenstates
$(\nu_1, \nu_2, \nu_3)$ to the neutrino flavor eigenstates
$(\nu_e, \nu_\mu, \nu_\tau)$. 

Different from the ``tri-maximal'' \cite{Cabibbo} or ``democratic''
(nearly bi-maximal) \cite{FX96} neutrino mixing scenarios,
the vanishing of $U_{e3}$ in the bi-maximal mixing scenario
assures the absence of $CP$ violation and
an exact decoupling between solar ($\nu_e \rightarrow \nu_\mu$)
and atmospheric ($\nu_\mu \rightarrow \nu_\tau$)
neutrino oscillations with identical 
mixing factors 
(i.e., $\sin^2 2\theta_{\rm sun} = \sin^2 2\theta_{\rm atm} =1$).
At present this special ansatz
seems to be more favored by the vacuum oscillation solution
than by the large-angle MSW solution to the solar neutrino problem.

In this short note I propose a simple extension of the naive
bi-maximal neutrino mixing pattern given above, so as to
accommodate possible large $CP$ violation in the lepton sector
and to allow much flexibility in accounting for the solar neutrino
problem. The new lepton flavor mixing matrix takes the following form
\footnote{Here only the Dirac-type $CP$ phase is taken
into account, as the Majorana-type ones have no effect in 
neutrino oscillations.}:
\begin{equation}
V \; =\; \left ( \matrix{
\frac{c}{\sqrt{2}}	& \frac{c}{\sqrt{2}}	& -is \cr\cr
-\frac{A}{2}	& \frac{A^*}{2}	& \frac{c}{\sqrt{2}} \cr\cr
\frac{A^*}{2}	& -\frac{A}{2}	& \frac{c}{\sqrt{2}} \cr} \right ) \;\; ,
\end{equation}
where $s \equiv \sin \theta$, $c \equiv \cos \theta$, and 
$A = 1 + is$. The mixing angle $\theta$ measures a slight coupling
between solar and atmospheric neutrino oscillations, whose
mass-squared differences are 
\begin{eqnarray}
\Delta m^2_{\rm sun} & = & \left |m^2_2 - m^2_1 \right | \;\; ,
\nonumber \\
\Delta m^2_{\rm atm} & = & \left |m^2_3 - m^2_2 \right | \;\; ,
\end{eqnarray}
respectively. The ``observed''
hierarchy $\Delta m^2_{\rm sun} \ll \Delta m^2_{\rm atm}$
represents our today's understanding of the neutrino mass spectrum
\footnote{Throughout this work we do not take the LSND evidence
for neutrino oscillations \cite{LSND}, which was not confirmed by
the KARMEN experiment \cite{KARMEN}, into account.}.

Clearly
$U$ can be reproduced from $V$ with $\theta =0$. 
The rephasing-invariant strength of $CP$ violation turns out to be
\begin{eqnarray}
{\cal J} & = & {\rm Im} \left (V_{i\alpha} V_{j\beta}
V^*_{i\beta} V^*_{j\alpha} \right ) \nonumber \\
& = & \frac{s ~ c^2}{4} \;\; ,
\end{eqnarray}
in which $(i, j)$ run over $(e, \mu)$, $(\mu, \tau)$ or $(\tau, e)$ and
$(\alpha, \beta)$ over $(1, 2)$, $(2, 3)$ or $(3, 1)$.
The mixing factors of solar ($\nu_e \rightarrow \nu_e$ disappearance)
and atmospheric ($\nu_\mu \rightarrow \nu_\mu$ disappearance)
neutrino oscillations read
\begin{eqnarray}
\sin^2 2\theta_{\rm sun} & = & c^4 \;\; , \nonumber \\
\sin^2 2\theta_{\rm atm} & = & 1- s^4 \;\; .
\end{eqnarray}
The correlation between ${\cal J}$ and $|V_{e3}|^2$ is illustrated
in Fig. 1(a) for $|V_{e3}|^2 \leq 0.1$, and that between 
$\sin^2 2\theta_{\rm sun}$ (or $\sin^2 2\theta_{\rm atm}$) and
$|V_{e3}|^2$ is shown in Fig. 2(b). Some comments are in order.
\begin{itemize}

\item	The strength of 
$CP$ violation in this new neutrino mixing scenario can
be as large as few percent. A signal of $CP$ (or $T$) violation
could be measured from the probability asymmetry between
$\nu_\mu \rightarrow \nu_e$ and $\bar{\nu}_\mu \rightarrow
\bar{\nu}_e$ (or $\nu_e \rightarrow \nu_\mu$) in the future
long-baseline neutrino experiments with $L \sim E/\Delta m^2_{\rm sun}$.
Such a measurment is feasible if and only if the large-angle
MSW oscillation is the true solution to the solar neutrino problem.

\item	The mixing factor of solar neutrino oscillations is 
sufficiently large, compatible with either the large-angle MSW solution
or the vacuum oscillation solution to the solar neutrino 
problem \cite{Petcov}.

\item	The mixing factor of atmospheric neutrino oscillations
is nearly maximal for reasonable values of $|V_{e3}|^2$. This 
result is particularly favored by the Super-Kamiokande \cite{SK} and CHOOZ
\cite{CHOOZ} experiments.
\end{itemize}

It is also worth pointing out that the lepton flavor mixing matrix $V$
is symmetric about its axis $V_{e3}$-$V_{\mu 2}$-$V_{\tau 1}$.
In comparison, the quark flavor mixing matrix (i.e., the CKM matrix)
is approximately
symmetric about its axis $V_{ud}$-$V_{cs}$-$V_{tb}$ \cite{Xing95}. 
This qualitative difference in the textures of quark and lepton
mixing matrices could have a yet unknown dynamical reason. 

There are of course some other possibilities, depending on the number
of free parameters to be introduced into $U$, towards a slight
modification of the exactly bi-maximal neutrino mixing pattern.
In Ref. \cite{Sumino}, for instance, $U$ has been modified to
$U' = R^{\rm T}_{12} (\theta_{12}) ~ U$, where $R_{12}(\theta_{12})$
describes a small rotation in the real (1,2) plane to diagonalize
the charged lepton mass matrix 
($\theta_{12} \approx \arcsin \sqrt{m_e/m_\mu} \approx 4^{\circ}$).
This scenario, in which $CP$ symmetry remains conserved,
predicts $|U'_{e3}|^2 \approx 0.002$,
$\sin^2 2\theta_{\rm sun} \approx 0.985$, and $\sin^2 2\theta_{\rm atm}
\approx 1.000$. It is therefore distinguishable from the new mixing
pattern $V$ proposed in Eq. (2), through the delicate detection of
leptonic $CP$ violation in the long-baseline neutrino experiments
or through the accurate determination of mixing factors in the atmospheric
and solar neutrino experiments. In Ref. \cite{Stech} the author
presented a different nearly bi-maximal neutrino mixing
ansatz, based on the arguments of lepton-quark similarities and the
assumed textures of charged lepton and neutrino mass matrices. 
Such an ansatz involves several free parameters, whose values
are completely unknown. Hence its
interpretation of current neutrino oscillation data relies
somehow on the fine-tuning of those parameters, and its consequences
on the long-baseline neutrino oscillations are not as unique
as those of the simpler and more instructive scenarios discussed above.

Finally I should emphasize that the bi-maximal mixing pattern $U$ 
in Eq. (1) and its revisited version $V$ in Eq. (2) 
require some special flavor symmetries
to be imposed on the charged lepton and neutrino mass matrices \cite{FX96}.
It is more likely that $V$ (or $U$) serves as the leading-order approximation
of a more complicated flavor mixing matrix. At present, however, such
a simplified scenario is very instructive and useful to explore the
main features of lepton flavor mixing and $CP$ violation through
neutrino oscillations.

\begin{figure}
\setlength{\unitlength}{0.240900pt}
\ifx\plotpoint\undefined\newsavebox{\plotpoint}\fi
\sbox{\plotpoint}{\rule[-0.200pt]{0.400pt}{0.400pt}}%
\begin{picture}(1200,1080)(-260,0)
\font\gnuplot=cmr10 at 10pt
\gnuplot
\sbox{\plotpoint}{\rule[-0.200pt]{0.400pt}{0.400pt}}%
\put(140.0,123.0){\rule[-0.200pt]{4.818pt}{0.400pt}}
\put(120,123){\makebox(0,0)[r]{$0\%$}}
\put(1159.0,123.0){\rule[-0.200pt]{4.818pt}{0.400pt}}
\put(140.0,238.0){\rule[-0.200pt]{4.818pt}{0.400pt}}
\put(1159.0,238.0){\rule[-0.200pt]{4.818pt}{0.400pt}}
\put(140.0,352.0){\rule[-0.200pt]{4.818pt}{0.400pt}}
\put(120,352){\makebox(0,0)[r]{$2\%$}}
\put(1159.0,352.0){\rule[-0.200pt]{4.818pt}{0.400pt}}
\put(140.0,467.0){\rule[-0.200pt]{4.818pt}{0.400pt}}
\put(1159.0,467.0){\rule[-0.200pt]{4.818pt}{0.400pt}}
\put(140.0,582.0){\rule[-0.200pt]{4.818pt}{0.400pt}}
\put(120,582){\makebox(0,0)[r]{$4\%$}}
\put(1159.0,582.0){\rule[-0.200pt]{4.818pt}{0.400pt}}
\put(140.0,696.0){\rule[-0.200pt]{4.818pt}{0.400pt}}
\put(1159.0,696.0){\rule[-0.200pt]{4.818pt}{0.400pt}}
\put(140.0,811.0){\rule[-0.200pt]{4.818pt}{0.400pt}}
\put(120,811){\makebox(0,0)[r]{$6\%$}}
\put(1159.0,811.0){\rule[-0.200pt]{4.818pt}{0.400pt}}
\put(140.0,925.0){\rule[-0.200pt]{4.818pt}{0.400pt}}
\put(1159.0,925.0){\rule[-0.200pt]{4.818pt}{0.400pt}}
\put(140.0,1040.0){\rule[-0.200pt]{4.818pt}{0.400pt}}
\put(120,1040){\makebox(0,0)[r]{$8\%$}}
\put(1159.0,1040.0){\rule[-0.200pt]{4.818pt}{0.400pt}}
\put(140.0,123.0){\rule[-0.200pt]{0.400pt}{4.818pt}}
\put(140,82){\makebox(0,0){$10^{-3}$}}
\put(140.0,1020.0){\rule[-0.200pt]{0.400pt}{4.818pt}}
\put(296.0,123.0){\rule[-0.200pt]{0.400pt}{2.409pt}}
\put(296.0,1030.0){\rule[-0.200pt]{0.400pt}{2.409pt}}
\put(388.0,123.0){\rule[-0.200pt]{0.400pt}{2.409pt}}
\put(388.0,1030.0){\rule[-0.200pt]{0.400pt}{2.409pt}}
\put(453.0,123.0){\rule[-0.200pt]{0.400pt}{2.409pt}}
\put(453.0,1030.0){\rule[-0.200pt]{0.400pt}{2.409pt}}
\put(503.0,123.0){\rule[-0.200pt]{0.400pt}{2.409pt}}
\put(503.0,1030.0){\rule[-0.200pt]{0.400pt}{2.409pt}}
\put(544.0,123.0){\rule[-0.200pt]{0.400pt}{2.409pt}}
\put(544.0,1030.0){\rule[-0.200pt]{0.400pt}{2.409pt}}
\put(579.0,123.0){\rule[-0.200pt]{0.400pt}{2.409pt}}
\put(579.0,1030.0){\rule[-0.200pt]{0.400pt}{2.409pt}}
\put(609.0,123.0){\rule[-0.200pt]{0.400pt}{2.409pt}}
\put(609.0,1030.0){\rule[-0.200pt]{0.400pt}{2.409pt}}
\put(636.0,123.0){\rule[-0.200pt]{0.400pt}{2.409pt}}
\put(636.0,1030.0){\rule[-0.200pt]{0.400pt}{2.409pt}}
\put(660.0,123.0){\rule[-0.200pt]{0.400pt}{4.818pt}}
\put(660,82){\makebox(0,0){$10^{-2}$}}
\put(660.0,1020.0){\rule[-0.200pt]{0.400pt}{4.818pt}}
\put(816.0,123.0){\rule[-0.200pt]{0.400pt}{2.409pt}}
\put(816.0,1030.0){\rule[-0.200pt]{0.400pt}{2.409pt}}
\put(907.0,123.0){\rule[-0.200pt]{0.400pt}{2.409pt}}
\put(907.0,1030.0){\rule[-0.200pt]{0.400pt}{2.409pt}}
\put(972.0,123.0){\rule[-0.200pt]{0.400pt}{2.409pt}}
\put(972.0,1030.0){\rule[-0.200pt]{0.400pt}{2.409pt}}
\put(1023.0,123.0){\rule[-0.200pt]{0.400pt}{2.409pt}}
\put(1023.0,1030.0){\rule[-0.200pt]{0.400pt}{2.409pt}}
\put(1064.0,123.0){\rule[-0.200pt]{0.400pt}{2.409pt}}
\put(1064.0,1030.0){\rule[-0.200pt]{0.400pt}{2.409pt}}
\put(1099.0,123.0){\rule[-0.200pt]{0.400pt}{2.409pt}}
\put(1099.0,1030.0){\rule[-0.200pt]{0.400pt}{2.409pt}}
\put(1129.0,123.0){\rule[-0.200pt]{0.400pt}{2.409pt}}
\put(1129.0,1030.0){\rule[-0.200pt]{0.400pt}{2.409pt}}
\put(1155.0,123.0){\rule[-0.200pt]{0.400pt}{2.409pt}}
\put(1155.0,1030.0){\rule[-0.200pt]{0.400pt}{2.409pt}}
\put(1179.0,123.0){\rule[-0.200pt]{0.400pt}{4.818pt}}
\put(1179,82){\makebox(0,0){$10^{-1}$}}
\put(1179.0,1020.0){\rule[-0.200pt]{0.400pt}{4.818pt}}
\put(140.0,123.0){\rule[-0.200pt]{250.295pt}{0.400pt}}
\put(1179.0,123.0){\rule[-0.200pt]{0.400pt}{220.905pt}}
\put(140.0,1040.0){\rule[-0.200pt]{250.295pt}{0.400pt}}
\put(659,0){\makebox(0,0){$|V_{e3}|^2$}}

\put(659,460){\makebox(0,0){$\cal J$}}

\put(230,970){\makebox(0,0){(a)}}

\put(140.0,123.0){\rule[-0.200pt]{0.400pt}{220.905pt}}
\put(140,214){\usebox{\plotpoint}}
\multiput(140.00,214.58)(2.123,0.498){71}{\rule{1.786pt}{0.120pt}}
\multiput(140.00,213.17)(152.292,37.000){2}{\rule{0.893pt}{0.400pt}}
\multiput(296.00,251.58)(1.656,0.497){53}{\rule{1.414pt}{0.120pt}}
\multiput(296.00,250.17)(89.065,28.000){2}{\rule{0.707pt}{0.400pt}}
\multiput(388.00,279.58)(1.310,0.497){47}{\rule{1.140pt}{0.120pt}}
\multiput(388.00,278.17)(62.634,25.000){2}{\rule{0.570pt}{0.400pt}}
\multiput(453.00,304.58)(1.201,0.496){39}{\rule{1.052pt}{0.119pt}}
\multiput(453.00,303.17)(47.816,21.000){2}{\rule{0.526pt}{0.400pt}}
\multiput(503.00,325.58)(1.088,0.495){35}{\rule{0.963pt}{0.119pt}}
\multiput(503.00,324.17)(39.001,19.000){2}{\rule{0.482pt}{0.400pt}}
\multiput(544.00,344.58)(1.039,0.495){31}{\rule{0.924pt}{0.119pt}}
\multiput(544.00,343.17)(33.083,17.000){2}{\rule{0.462pt}{0.400pt}}
\multiput(579.00,361.58)(0.945,0.494){29}{\rule{0.850pt}{0.119pt}}
\multiput(579.00,360.17)(28.236,16.000){2}{\rule{0.425pt}{0.400pt}}
\multiput(609.00,377.58)(0.908,0.494){27}{\rule{0.820pt}{0.119pt}}
\multiput(609.00,376.17)(25.298,15.000){2}{\rule{0.410pt}{0.400pt}}
\multiput(636.00,392.58)(0.805,0.494){27}{\rule{0.740pt}{0.119pt}}
\multiput(636.00,391.17)(22.464,15.000){2}{\rule{0.370pt}{0.400pt}}
\multiput(660.00,407.58)(0.814,0.493){23}{\rule{0.746pt}{0.119pt}}
\multiput(660.00,406.17)(19.451,13.000){2}{\rule{0.373pt}{0.400pt}}
\multiput(681.00,420.58)(0.774,0.493){23}{\rule{0.715pt}{0.119pt}}
\multiput(681.00,419.17)(18.515,13.000){2}{\rule{0.358pt}{0.400pt}}
\multiput(701.00,433.58)(0.755,0.492){21}{\rule{0.700pt}{0.119pt}}
\multiput(701.00,432.17)(16.547,12.000){2}{\rule{0.350pt}{0.400pt}}
\multiput(719.00,445.58)(0.669,0.492){21}{\rule{0.633pt}{0.119pt}}
\multiput(719.00,444.17)(14.685,12.000){2}{\rule{0.317pt}{0.400pt}}
\multiput(735.00,457.58)(0.669,0.492){21}{\rule{0.633pt}{0.119pt}}
\multiput(735.00,456.17)(14.685,12.000){2}{\rule{0.317pt}{0.400pt}}
\multiput(751.00,469.58)(0.684,0.492){19}{\rule{0.645pt}{0.118pt}}
\multiput(751.00,468.17)(13.660,11.000){2}{\rule{0.323pt}{0.400pt}}
\multiput(766.00,480.58)(0.652,0.491){17}{\rule{0.620pt}{0.118pt}}
\multiput(766.00,479.17)(11.713,10.000){2}{\rule{0.310pt}{0.400pt}}
\multiput(779.00,490.58)(0.590,0.492){19}{\rule{0.573pt}{0.118pt}}
\multiput(779.00,489.17)(11.811,11.000){2}{\rule{0.286pt}{0.400pt}}
\multiput(792.00,501.59)(0.669,0.489){15}{\rule{0.633pt}{0.118pt}}
\multiput(792.00,500.17)(10.685,9.000){2}{\rule{0.317pt}{0.400pt}}
\multiput(804.00,510.58)(0.600,0.491){17}{\rule{0.580pt}{0.118pt}}
\multiput(804.00,509.17)(10.796,10.000){2}{\rule{0.290pt}{0.400pt}}
\multiput(816.00,520.58)(0.547,0.491){17}{\rule{0.540pt}{0.118pt}}
\multiput(816.00,519.17)(9.879,10.000){2}{\rule{0.270pt}{0.400pt}}
\multiput(827.00,530.59)(0.553,0.489){15}{\rule{0.544pt}{0.118pt}}
\multiput(827.00,529.17)(8.870,9.000){2}{\rule{0.272pt}{0.400pt}}
\multiput(837.00,539.59)(0.553,0.489){15}{\rule{0.544pt}{0.118pt}}
\multiput(837.00,538.17)(8.870,9.000){2}{\rule{0.272pt}{0.400pt}}
\multiput(847.00,548.59)(0.626,0.488){13}{\rule{0.600pt}{0.117pt}}
\multiput(847.00,547.17)(8.755,8.000){2}{\rule{0.300pt}{0.400pt}}
\multiput(857.00,556.59)(0.495,0.489){15}{\rule{0.500pt}{0.118pt}}
\multiput(857.00,555.17)(7.962,9.000){2}{\rule{0.250pt}{0.400pt}}
\multiput(866.00,565.59)(0.560,0.488){13}{\rule{0.550pt}{0.117pt}}
\multiput(866.00,564.17)(7.858,8.000){2}{\rule{0.275pt}{0.400pt}}
\multiput(875.00,573.59)(0.560,0.488){13}{\rule{0.550pt}{0.117pt}}
\multiput(875.00,572.17)(7.858,8.000){2}{\rule{0.275pt}{0.400pt}}
\multiput(884.00,581.59)(0.494,0.488){13}{\rule{0.500pt}{0.117pt}}
\multiput(884.00,580.17)(6.962,8.000){2}{\rule{0.250pt}{0.400pt}}
\multiput(892.00,589.59)(0.494,0.488){13}{\rule{0.500pt}{0.117pt}}
\multiput(892.00,588.17)(6.962,8.000){2}{\rule{0.250pt}{0.400pt}}
\multiput(900.00,597.59)(0.492,0.485){11}{\rule{0.500pt}{0.117pt}}
\multiput(900.00,596.17)(5.962,7.000){2}{\rule{0.250pt}{0.400pt}}
\multiput(907.00,604.59)(0.494,0.488){13}{\rule{0.500pt}{0.117pt}}
\multiput(907.00,603.17)(6.962,8.000){2}{\rule{0.250pt}{0.400pt}}
\multiput(915.00,612.59)(0.492,0.485){11}{\rule{0.500pt}{0.117pt}}
\multiput(915.00,611.17)(5.962,7.000){2}{\rule{0.250pt}{0.400pt}}
\multiput(922.00,619.59)(0.492,0.485){11}{\rule{0.500pt}{0.117pt}}
\multiput(922.00,618.17)(5.962,7.000){2}{\rule{0.250pt}{0.400pt}}
\multiput(929.00,626.59)(0.492,0.485){11}{\rule{0.500pt}{0.117pt}}
\multiput(929.00,625.17)(5.962,7.000){2}{\rule{0.250pt}{0.400pt}}
\multiput(936.59,633.00)(0.482,0.581){9}{\rule{0.116pt}{0.567pt}}
\multiput(935.17,633.00)(6.000,5.824){2}{\rule{0.400pt}{0.283pt}}
\multiput(942.59,640.00)(0.482,0.581){9}{\rule{0.116pt}{0.567pt}}
\multiput(941.17,640.00)(6.000,5.824){2}{\rule{0.400pt}{0.283pt}}
\multiput(948.00,647.59)(0.492,0.485){11}{\rule{0.500pt}{0.117pt}}
\multiput(948.00,646.17)(5.962,7.000){2}{\rule{0.250pt}{0.400pt}}
\multiput(955.00,654.59)(0.491,0.482){9}{\rule{0.500pt}{0.116pt}}
\multiput(955.00,653.17)(4.962,6.000){2}{\rule{0.250pt}{0.400pt}}
\multiput(961.59,660.00)(0.482,0.581){9}{\rule{0.116pt}{0.567pt}}
\multiput(960.17,660.00)(6.000,5.824){2}{\rule{0.400pt}{0.283pt}}
\multiput(967.59,667.00)(0.477,0.599){7}{\rule{0.115pt}{0.580pt}}
\multiput(966.17,667.00)(5.000,4.796){2}{\rule{0.400pt}{0.290pt}}
\multiput(972.00,673.59)(0.491,0.482){9}{\rule{0.500pt}{0.116pt}}
\multiput(972.00,672.17)(4.962,6.000){2}{\rule{0.250pt}{0.400pt}}
\multiput(978.59,679.00)(0.477,0.710){7}{\rule{0.115pt}{0.660pt}}
\multiput(977.17,679.00)(5.000,5.630){2}{\rule{0.400pt}{0.330pt}}
\multiput(983.00,686.59)(0.491,0.482){9}{\rule{0.500pt}{0.116pt}}
\multiput(983.00,685.17)(4.962,6.000){2}{\rule{0.250pt}{0.400pt}}
\multiput(989.59,692.00)(0.477,0.599){7}{\rule{0.115pt}{0.580pt}}
\multiput(988.17,692.00)(5.000,4.796){2}{\rule{0.400pt}{0.290pt}}
\multiput(994.59,698.00)(0.477,0.599){7}{\rule{0.115pt}{0.580pt}}
\multiput(993.17,698.00)(5.000,4.796){2}{\rule{0.400pt}{0.290pt}}
\multiput(999.00,704.59)(0.487,0.477){7}{\rule{0.500pt}{0.115pt}}
\multiput(999.00,703.17)(3.962,5.000){2}{\rule{0.250pt}{0.400pt}}
\multiput(1004.59,709.00)(0.477,0.599){7}{\rule{0.115pt}{0.580pt}}
\multiput(1003.17,709.00)(5.000,4.796){2}{\rule{0.400pt}{0.290pt}}
\multiput(1009.60,715.00)(0.468,0.774){5}{\rule{0.113pt}{0.700pt}}
\multiput(1008.17,715.00)(4.000,4.547){2}{\rule{0.400pt}{0.350pt}}
\multiput(1013.00,721.59)(0.487,0.477){7}{\rule{0.500pt}{0.115pt}}
\multiput(1013.00,720.17)(3.962,5.000){2}{\rule{0.250pt}{0.400pt}}
\multiput(1018.59,726.00)(0.477,0.599){7}{\rule{0.115pt}{0.580pt}}
\multiput(1017.17,726.00)(5.000,4.796){2}{\rule{0.400pt}{0.290pt}}
\multiput(1023.60,732.00)(0.468,0.627){5}{\rule{0.113pt}{0.600pt}}
\multiput(1022.17,732.00)(4.000,3.755){2}{\rule{0.400pt}{0.300pt}}
\multiput(1027.60,737.00)(0.468,0.627){5}{\rule{0.113pt}{0.600pt}}
\multiput(1026.17,737.00)(4.000,3.755){2}{\rule{0.400pt}{0.300pt}}
\multiput(1031.59,742.00)(0.477,0.599){7}{\rule{0.115pt}{0.580pt}}
\multiput(1030.17,742.00)(5.000,4.796){2}{\rule{0.400pt}{0.290pt}}
\multiput(1036.60,748.00)(0.468,0.627){5}{\rule{0.113pt}{0.600pt}}
\multiput(1035.17,748.00)(4.000,3.755){2}{\rule{0.400pt}{0.300pt}}
\multiput(1040.60,753.00)(0.468,0.627){5}{\rule{0.113pt}{0.600pt}}
\multiput(1039.17,753.00)(4.000,3.755){2}{\rule{0.400pt}{0.300pt}}
\multiput(1044.60,758.00)(0.468,0.627){5}{\rule{0.113pt}{0.600pt}}
\multiput(1043.17,758.00)(4.000,3.755){2}{\rule{0.400pt}{0.300pt}}
\multiput(1048.60,763.00)(0.468,0.627){5}{\rule{0.113pt}{0.600pt}}
\multiput(1047.17,763.00)(4.000,3.755){2}{\rule{0.400pt}{0.300pt}}
\multiput(1052.60,768.00)(0.468,0.627){5}{\rule{0.113pt}{0.600pt}}
\multiput(1051.17,768.00)(4.000,3.755){2}{\rule{0.400pt}{0.300pt}}
\multiput(1056.60,773.00)(0.468,0.627){5}{\rule{0.113pt}{0.600pt}}
\multiput(1055.17,773.00)(4.000,3.755){2}{\rule{0.400pt}{0.300pt}}
\multiput(1060.60,778.00)(0.468,0.627){5}{\rule{0.113pt}{0.600pt}}
\multiput(1059.17,778.00)(4.000,3.755){2}{\rule{0.400pt}{0.300pt}}
\multiput(1064.61,783.00)(0.447,0.909){3}{\rule{0.108pt}{0.767pt}}
\multiput(1063.17,783.00)(3.000,3.409){2}{\rule{0.400pt}{0.383pt}}
\multiput(1067.00,788.60)(0.481,0.468){5}{\rule{0.500pt}{0.113pt}}
\multiput(1067.00,787.17)(2.962,4.000){2}{\rule{0.250pt}{0.400pt}}
\multiput(1071.60,792.00)(0.468,0.627){5}{\rule{0.113pt}{0.600pt}}
\multiput(1070.17,792.00)(4.000,3.755){2}{\rule{0.400pt}{0.300pt}}
\multiput(1075.61,797.00)(0.447,0.909){3}{\rule{0.108pt}{0.767pt}}
\multiput(1074.17,797.00)(3.000,3.409){2}{\rule{0.400pt}{0.383pt}}
\multiput(1078.00,802.60)(0.481,0.468){5}{\rule{0.500pt}{0.113pt}}
\multiput(1078.00,801.17)(2.962,4.000){2}{\rule{0.250pt}{0.400pt}}
\multiput(1082.61,806.00)(0.447,0.909){3}{\rule{0.108pt}{0.767pt}}
\multiput(1081.17,806.00)(3.000,3.409){2}{\rule{0.400pt}{0.383pt}}
\multiput(1085.00,811.60)(0.481,0.468){5}{\rule{0.500pt}{0.113pt}}
\multiput(1085.00,810.17)(2.962,4.000){2}{\rule{0.250pt}{0.400pt}}
\multiput(1089.61,815.00)(0.447,0.685){3}{\rule{0.108pt}{0.633pt}}
\multiput(1088.17,815.00)(3.000,2.685){2}{\rule{0.400pt}{0.317pt}}
\multiput(1092.61,819.00)(0.447,0.909){3}{\rule{0.108pt}{0.767pt}}
\multiput(1091.17,819.00)(3.000,3.409){2}{\rule{0.400pt}{0.383pt}}
\multiput(1095.00,824.60)(0.481,0.468){5}{\rule{0.500pt}{0.113pt}}
\multiput(1095.00,823.17)(2.962,4.000){2}{\rule{0.250pt}{0.400pt}}
\multiput(1099.61,828.00)(0.447,0.685){3}{\rule{0.108pt}{0.633pt}}
\multiput(1098.17,828.00)(3.000,2.685){2}{\rule{0.400pt}{0.317pt}}
\multiput(1102.61,832.00)(0.447,0.909){3}{\rule{0.108pt}{0.767pt}}
\multiput(1101.17,832.00)(3.000,3.409){2}{\rule{0.400pt}{0.383pt}}
\multiput(1105.61,837.00)(0.447,0.685){3}{\rule{0.108pt}{0.633pt}}
\multiput(1104.17,837.00)(3.000,2.685){2}{\rule{0.400pt}{0.317pt}}
\multiput(1108.61,841.00)(0.447,0.685){3}{\rule{0.108pt}{0.633pt}}
\multiput(1107.17,841.00)(3.000,2.685){2}{\rule{0.400pt}{0.317pt}}
\multiput(1111.61,845.00)(0.447,0.685){3}{\rule{0.108pt}{0.633pt}}
\multiput(1110.17,845.00)(3.000,2.685){2}{\rule{0.400pt}{0.317pt}}
\multiput(1114.61,849.00)(0.447,0.685){3}{\rule{0.108pt}{0.633pt}}
\multiput(1113.17,849.00)(3.000,2.685){2}{\rule{0.400pt}{0.317pt}}
\multiput(1117.61,853.00)(0.447,0.685){3}{\rule{0.108pt}{0.633pt}}
\multiput(1116.17,853.00)(3.000,2.685){2}{\rule{0.400pt}{0.317pt}}
\multiput(1120.61,857.00)(0.447,0.685){3}{\rule{0.108pt}{0.633pt}}
\multiput(1119.17,857.00)(3.000,2.685){2}{\rule{0.400pt}{0.317pt}}
\multiput(1123.61,861.00)(0.447,0.685){3}{\rule{0.108pt}{0.633pt}}
\multiput(1122.17,861.00)(3.000,2.685){2}{\rule{0.400pt}{0.317pt}}
\multiput(1126.61,865.00)(0.447,0.685){3}{\rule{0.108pt}{0.633pt}}
\multiput(1125.17,865.00)(3.000,2.685){2}{\rule{0.400pt}{0.317pt}}
\put(1129.17,869){\rule{0.400pt}{0.900pt}}
\multiput(1128.17,869.00)(2.000,2.132){2}{\rule{0.400pt}{0.450pt}}
\multiput(1131.00,873.61)(0.462,0.447){3}{\rule{0.500pt}{0.108pt}}
\multiput(1131.00,872.17)(1.962,3.000){2}{\rule{0.250pt}{0.400pt}}
\multiput(1134.61,876.00)(0.447,0.685){3}{\rule{0.108pt}{0.633pt}}
\multiput(1133.17,876.00)(3.000,2.685){2}{\rule{0.400pt}{0.317pt}}
\multiput(1137.61,880.00)(0.447,0.685){3}{\rule{0.108pt}{0.633pt}}
\multiput(1136.17,880.00)(3.000,2.685){2}{\rule{0.400pt}{0.317pt}}
\put(1140.17,884){\rule{0.400pt}{0.700pt}}
\multiput(1139.17,884.00)(2.000,1.547){2}{\rule{0.400pt}{0.350pt}}
\multiput(1142.61,887.00)(0.447,0.685){3}{\rule{0.108pt}{0.633pt}}
\multiput(1141.17,887.00)(3.000,2.685){2}{\rule{0.400pt}{0.317pt}}
\multiput(1145.61,891.00)(0.447,0.685){3}{\rule{0.108pt}{0.633pt}}
\multiput(1144.17,891.00)(3.000,2.685){2}{\rule{0.400pt}{0.317pt}}
\put(1148.17,895){\rule{0.400pt}{0.700pt}}
\multiput(1147.17,895.00)(2.000,1.547){2}{\rule{0.400pt}{0.350pt}}
\multiput(1150.61,898.00)(0.447,0.685){3}{\rule{0.108pt}{0.633pt}}
\multiput(1149.17,898.00)(3.000,2.685){2}{\rule{0.400pt}{0.317pt}}
\put(1153.17,902){\rule{0.400pt}{0.700pt}}
\multiput(1152.17,902.00)(2.000,1.547){2}{\rule{0.400pt}{0.350pt}}
\multiput(1155.61,905.00)(0.447,0.685){3}{\rule{0.108pt}{0.633pt}}
\multiput(1154.17,905.00)(3.000,2.685){2}{\rule{0.400pt}{0.317pt}}
\put(1158.17,909){\rule{0.400pt}{0.700pt}}
\multiput(1157.17,909.00)(2.000,1.547){2}{\rule{0.400pt}{0.350pt}}
\multiput(1160.61,912.00)(0.447,0.685){3}{\rule{0.108pt}{0.633pt}}
\multiput(1159.17,912.00)(3.000,2.685){2}{\rule{0.400pt}{0.317pt}}
\put(1163.17,916){\rule{0.400pt}{0.700pt}}
\multiput(1162.17,916.00)(2.000,1.547){2}{\rule{0.400pt}{0.350pt}}
\put(1165.17,919){\rule{0.400pt}{0.700pt}}
\multiput(1164.17,919.00)(2.000,1.547){2}{\rule{0.400pt}{0.350pt}}
\multiput(1167.61,922.00)(0.447,0.685){3}{\rule{0.108pt}{0.633pt}}
\multiput(1166.17,922.00)(3.000,2.685){2}{\rule{0.400pt}{0.317pt}}
\put(1170.17,926){\rule{0.400pt}{0.700pt}}
\multiput(1169.17,926.00)(2.000,1.547){2}{\rule{0.400pt}{0.350pt}}
\put(1172.17,929){\rule{0.400pt}{0.700pt}}
\multiput(1171.17,929.00)(2.000,1.547){2}{\rule{0.400pt}{0.350pt}}
\multiput(1174.00,932.61)(0.462,0.447){3}{\rule{0.500pt}{0.108pt}}
\multiput(1174.00,931.17)(1.962,3.000){2}{\rule{0.250pt}{0.400pt}}
\end{picture}

\vspace{1.1cm}

\setlength{\unitlength}{0.240900pt}
\ifx\plotpoint\undefined\newsavebox{\plotpoint}\fi
\sbox{\plotpoint}{\rule[-0.200pt]{0.400pt}{0.400pt}}%
\begin{picture}(1200,1080)(-260,0)
\font\gnuplot=cmr10 at 10pt
\gnuplot
\sbox{\plotpoint}{\rule[-0.200pt]{0.400pt}{0.400pt}}%
\put(140.0,123.0){\rule[-0.200pt]{4.818pt}{0.400pt}}
\put(120,123){\makebox(0,0)[r]{0.8}}
\put(1159.0,123.0){\rule[-0.200pt]{4.818pt}{0.400pt}}
\put(140.0,215.0){\rule[-0.200pt]{4.818pt}{0.400pt}}
\put(1159.0,215.0){\rule[-0.200pt]{4.818pt}{0.400pt}}
\put(140.0,306.0){\rule[-0.200pt]{4.818pt}{0.400pt}}
\put(1159.0,306.0){\rule[-0.200pt]{4.818pt}{0.400pt}}
\put(140.0,398.0){\rule[-0.200pt]{4.818pt}{0.400pt}}
\put(1159.0,398.0){\rule[-0.200pt]{4.818pt}{0.400pt}}
\put(140.0,490.0){\rule[-0.200pt]{4.818pt}{0.400pt}}
\put(1159.0,490.0){\rule[-0.200pt]{4.818pt}{0.400pt}}
\put(140.0,582.0){\rule[-0.200pt]{4.818pt}{0.400pt}}
\put(120,582){\makebox(0,0)[r]{0.9}}
\put(1159.0,582.0){\rule[-0.200pt]{4.818pt}{0.400pt}}
\put(140.0,673.0){\rule[-0.200pt]{4.818pt}{0.400pt}}
\put(1159.0,673.0){\rule[-0.200pt]{4.818pt}{0.400pt}}
\put(140.0,765.0){\rule[-0.200pt]{4.818pt}{0.400pt}}
\put(1159.0,765.0){\rule[-0.200pt]{4.818pt}{0.400pt}}
\put(140.0,857.0){\rule[-0.200pt]{4.818pt}{0.400pt}}
\put(1159.0,857.0){\rule[-0.200pt]{4.818pt}{0.400pt}}
\put(140.0,948.0){\rule[-0.200pt]{4.818pt}{0.400pt}}
\put(1159.0,948.0){\rule[-0.200pt]{4.818pt}{0.400pt}}
\put(140.0,1040.0){\rule[-0.200pt]{4.818pt}{0.400pt}}
\put(120,1040){\makebox(0,0)[r]{1.0}}
\put(1159.0,1040.0){\rule[-0.200pt]{4.818pt}{0.400pt}}
\put(140.0,123.0){\rule[-0.200pt]{0.400pt}{4.818pt}}
\put(140,82){\makebox(0,0){$10^{-3}$}}
\put(140.0,1020.0){\rule[-0.200pt]{0.400pt}{4.818pt}}
\put(296.0,123.0){\rule[-0.200pt]{0.400pt}{2.409pt}}
\put(296.0,1030.0){\rule[-0.200pt]{0.400pt}{2.409pt}}
\put(388.0,123.0){\rule[-0.200pt]{0.400pt}{2.409pt}}
\put(388.0,1030.0){\rule[-0.200pt]{0.400pt}{2.409pt}}
\put(453.0,123.0){\rule[-0.200pt]{0.400pt}{2.409pt}}
\put(453.0,1030.0){\rule[-0.200pt]{0.400pt}{2.409pt}}
\put(503.0,123.0){\rule[-0.200pt]{0.400pt}{2.409pt}}
\put(503.0,1030.0){\rule[-0.200pt]{0.400pt}{2.409pt}}
\put(544.0,123.0){\rule[-0.200pt]{0.400pt}{2.409pt}}
\put(544.0,1030.0){\rule[-0.200pt]{0.400pt}{2.409pt}}
\put(579.0,123.0){\rule[-0.200pt]{0.400pt}{2.409pt}}
\put(579.0,1030.0){\rule[-0.200pt]{0.400pt}{2.409pt}}
\put(609.0,123.0){\rule[-0.200pt]{0.400pt}{2.409pt}}
\put(609.0,1030.0){\rule[-0.200pt]{0.400pt}{2.409pt}}
\put(636.0,123.0){\rule[-0.200pt]{0.400pt}{2.409pt}}
\put(636.0,1030.0){\rule[-0.200pt]{0.400pt}{2.409pt}}
\put(660.0,123.0){\rule[-0.200pt]{0.400pt}{4.818pt}}
\put(660,82){\makebox(0,0){$10^{-2}$}}
\put(660.0,1020.0){\rule[-0.200pt]{0.400pt}{4.818pt}}
\put(816.0,123.0){\rule[-0.200pt]{0.400pt}{2.409pt}}
\put(816.0,1030.0){\rule[-0.200pt]{0.400pt}{2.409pt}}
\put(907.0,123.0){\rule[-0.200pt]{0.400pt}{2.409pt}}
\put(907.0,1030.0){\rule[-0.200pt]{0.400pt}{2.409pt}}
\put(972.0,123.0){\rule[-0.200pt]{0.400pt}{2.409pt}}
\put(972.0,1030.0){\rule[-0.200pt]{0.400pt}{2.409pt}}
\put(1023.0,123.0){\rule[-0.200pt]{0.400pt}{2.409pt}}
\put(1023.0,1030.0){\rule[-0.200pt]{0.400pt}{2.409pt}}
\put(1064.0,123.0){\rule[-0.200pt]{0.400pt}{2.409pt}}
\put(1064.0,1030.0){\rule[-0.200pt]{0.400pt}{2.409pt}}
\put(1099.0,123.0){\rule[-0.200pt]{0.400pt}{2.409pt}}
\put(1099.0,1030.0){\rule[-0.200pt]{0.400pt}{2.409pt}}
\put(1129.0,123.0){\rule[-0.200pt]{0.400pt}{2.409pt}}
\put(1129.0,1030.0){\rule[-0.200pt]{0.400pt}{2.409pt}}
\put(1155.0,123.0){\rule[-0.200pt]{0.400pt}{2.409pt}}
\put(1155.0,1030.0){\rule[-0.200pt]{0.400pt}{2.409pt}}
\put(1179.0,123.0){\rule[-0.200pt]{0.400pt}{4.818pt}}
\put(1179,82){\makebox(0,0){$10^{-1}$}}
\put(1179.0,1020.0){\rule[-0.200pt]{0.400pt}{4.818pt}}
\put(140.0,123.0){\rule[-0.200pt]{250.295pt}{0.400pt}}
\put(1179.0,123.0){\rule[-0.200pt]{0.400pt}{220.905pt}}
\put(140.0,1040.0){\rule[-0.200pt]{250.295pt}{0.400pt}}
\put(659,0){\makebox(0,0){$|V_{e3}|^2$}}

\put(1030,980){\makebox(0,0){$\sin^2 2\theta_{\rm atm}$}}

\put(950,500){\makebox(0,0){$\sin^2 2\theta_{\rm sun}$}}

\put(230,200){\makebox(0,0){(b)}}

\put(140.0,123.0){\rule[-0.200pt]{0.400pt}{220.905pt}}
\put(140,1031){\usebox{\plotpoint}}
\multiput(140.00,1029.93)(9.057,-0.489){15}{\rule{7.033pt}{0.118pt}}
\multiput(140.00,1030.17)(141.402,-9.000){2}{\rule{3.517pt}{0.400pt}}
\multiput(296.00,1020.93)(5.329,-0.489){15}{\rule{4.189pt}{0.118pt}}
\multiput(296.00,1021.17)(83.306,-9.000){2}{\rule{2.094pt}{0.400pt}}
\multiput(388.00,1011.92)(3.362,-0.491){17}{\rule{2.700pt}{0.118pt}}
\multiput(388.00,1012.17)(59.396,-10.000){2}{\rule{1.350pt}{0.400pt}}
\multiput(453.00,1001.93)(2.883,-0.489){15}{\rule{2.322pt}{0.118pt}}
\multiput(453.00,1002.17)(45.180,-9.000){2}{\rule{1.161pt}{0.400pt}}
\multiput(503.00,992.93)(2.359,-0.489){15}{\rule{1.922pt}{0.118pt}}
\multiput(503.00,993.17)(37.010,-9.000){2}{\rule{0.961pt}{0.400pt}}
\multiput(544.00,983.93)(2.009,-0.489){15}{\rule{1.656pt}{0.118pt}}
\multiput(544.00,984.17)(31.564,-9.000){2}{\rule{0.828pt}{0.400pt}}
\multiput(579.00,974.93)(1.718,-0.489){15}{\rule{1.433pt}{0.118pt}}
\multiput(579.00,975.17)(27.025,-9.000){2}{\rule{0.717pt}{0.400pt}}
\multiput(609.00,965.93)(1.543,-0.489){15}{\rule{1.300pt}{0.118pt}}
\multiput(609.00,966.17)(24.302,-9.000){2}{\rule{0.650pt}{0.400pt}}
\multiput(636.00,956.93)(1.368,-0.489){15}{\rule{1.167pt}{0.118pt}}
\multiput(636.00,957.17)(21.579,-9.000){2}{\rule{0.583pt}{0.400pt}}
\multiput(660.00,947.93)(1.194,-0.489){15}{\rule{1.033pt}{0.118pt}}
\multiput(660.00,948.17)(18.855,-9.000){2}{\rule{0.517pt}{0.400pt}}
\multiput(681.00,938.93)(1.135,-0.489){15}{\rule{0.989pt}{0.118pt}}
\multiput(681.00,939.17)(17.948,-9.000){2}{\rule{0.494pt}{0.400pt}}
\multiput(701.00,929.93)(1.019,-0.489){15}{\rule{0.900pt}{0.118pt}}
\multiput(701.00,930.17)(16.132,-9.000){2}{\rule{0.450pt}{0.400pt}}
\multiput(719.00,920.93)(0.902,-0.489){15}{\rule{0.811pt}{0.118pt}}
\multiput(719.00,921.17)(14.316,-9.000){2}{\rule{0.406pt}{0.400pt}}
\multiput(735.00,911.92)(0.808,-0.491){17}{\rule{0.740pt}{0.118pt}}
\multiput(735.00,912.17)(14.464,-10.000){2}{\rule{0.370pt}{0.400pt}}
\multiput(751.00,901.93)(0.844,-0.489){15}{\rule{0.767pt}{0.118pt}}
\multiput(751.00,902.17)(13.409,-9.000){2}{\rule{0.383pt}{0.400pt}}
\multiput(766.00,892.93)(0.728,-0.489){15}{\rule{0.678pt}{0.118pt}}
\multiput(766.00,893.17)(11.593,-9.000){2}{\rule{0.339pt}{0.400pt}}
\multiput(779.00,883.93)(0.728,-0.489){15}{\rule{0.678pt}{0.118pt}}
\multiput(779.00,884.17)(11.593,-9.000){2}{\rule{0.339pt}{0.400pt}}
\multiput(792.00,874.93)(0.669,-0.489){15}{\rule{0.633pt}{0.118pt}}
\multiput(792.00,875.17)(10.685,-9.000){2}{\rule{0.317pt}{0.400pt}}
\multiput(804.00,865.93)(0.669,-0.489){15}{\rule{0.633pt}{0.118pt}}
\multiput(804.00,866.17)(10.685,-9.000){2}{\rule{0.317pt}{0.400pt}}
\multiput(816.00,856.93)(0.611,-0.489){15}{\rule{0.589pt}{0.118pt}}
\multiput(816.00,857.17)(9.778,-9.000){2}{\rule{0.294pt}{0.400pt}}
\multiput(827.00,847.93)(0.553,-0.489){15}{\rule{0.544pt}{0.118pt}}
\multiput(827.00,848.17)(8.870,-9.000){2}{\rule{0.272pt}{0.400pt}}
\multiput(837.00,838.93)(0.626,-0.488){13}{\rule{0.600pt}{0.117pt}}
\multiput(837.00,839.17)(8.755,-8.000){2}{\rule{0.300pt}{0.400pt}}
\multiput(847.00,830.93)(0.553,-0.489){15}{\rule{0.544pt}{0.118pt}}
\multiput(847.00,831.17)(8.870,-9.000){2}{\rule{0.272pt}{0.400pt}}
\multiput(857.00,821.93)(0.495,-0.489){15}{\rule{0.500pt}{0.118pt}}
\multiput(857.00,822.17)(7.962,-9.000){2}{\rule{0.250pt}{0.400pt}}
\multiput(866.00,812.93)(0.495,-0.489){15}{\rule{0.500pt}{0.118pt}}
\multiput(866.00,813.17)(7.962,-9.000){2}{\rule{0.250pt}{0.400pt}}
\multiput(875.00,803.93)(0.495,-0.489){15}{\rule{0.500pt}{0.118pt}}
\multiput(875.00,804.17)(7.962,-9.000){2}{\rule{0.250pt}{0.400pt}}
\multiput(884.59,793.72)(0.488,-0.560){13}{\rule{0.117pt}{0.550pt}}
\multiput(883.17,794.86)(8.000,-7.858){2}{\rule{0.400pt}{0.275pt}}
\multiput(892.59,784.72)(0.488,-0.560){13}{\rule{0.117pt}{0.550pt}}
\multiput(891.17,785.86)(8.000,-7.858){2}{\rule{0.400pt}{0.275pt}}
\multiput(900.59,775.45)(0.485,-0.645){11}{\rule{0.117pt}{0.614pt}}
\multiput(899.17,776.73)(7.000,-7.725){2}{\rule{0.400pt}{0.307pt}}
\multiput(907.59,766.72)(0.488,-0.560){13}{\rule{0.117pt}{0.550pt}}
\multiput(906.17,767.86)(8.000,-7.858){2}{\rule{0.400pt}{0.275pt}}
\multiput(915.59,757.45)(0.485,-0.645){11}{\rule{0.117pt}{0.614pt}}
\multiput(914.17,758.73)(7.000,-7.725){2}{\rule{0.400pt}{0.307pt}}
\multiput(922.59,748.45)(0.485,-0.645){11}{\rule{0.117pt}{0.614pt}}
\multiput(921.17,749.73)(7.000,-7.725){2}{\rule{0.400pt}{0.307pt}}
\multiput(929.59,739.69)(0.485,-0.569){11}{\rule{0.117pt}{0.557pt}}
\multiput(928.17,740.84)(7.000,-6.844){2}{\rule{0.400pt}{0.279pt}}
\multiput(936.59,731.09)(0.482,-0.762){9}{\rule{0.116pt}{0.700pt}}
\multiput(935.17,732.55)(6.000,-7.547){2}{\rule{0.400pt}{0.350pt}}
\multiput(942.59,722.09)(0.482,-0.762){9}{\rule{0.116pt}{0.700pt}}
\multiput(941.17,723.55)(6.000,-7.547){2}{\rule{0.400pt}{0.350pt}}
\multiput(948.59,713.45)(0.485,-0.645){11}{\rule{0.117pt}{0.614pt}}
\multiput(947.17,714.73)(7.000,-7.725){2}{\rule{0.400pt}{0.307pt}}
\multiput(955.59,704.09)(0.482,-0.762){9}{\rule{0.116pt}{0.700pt}}
\multiput(954.17,705.55)(6.000,-7.547){2}{\rule{0.400pt}{0.350pt}}
\multiput(961.59,695.09)(0.482,-0.762){9}{\rule{0.116pt}{0.700pt}}
\multiput(960.17,696.55)(6.000,-7.547){2}{\rule{0.400pt}{0.350pt}}
\multiput(967.59,685.93)(0.477,-0.821){7}{\rule{0.115pt}{0.740pt}}
\multiput(966.17,687.46)(5.000,-6.464){2}{\rule{0.400pt}{0.370pt}}
\multiput(972.59,678.09)(0.482,-0.762){9}{\rule{0.116pt}{0.700pt}}
\multiput(971.17,679.55)(6.000,-7.547){2}{\rule{0.400pt}{0.350pt}}
\multiput(978.59,668.60)(0.477,-0.933){7}{\rule{0.115pt}{0.820pt}}
\multiput(977.17,670.30)(5.000,-7.298){2}{\rule{0.400pt}{0.410pt}}
\multiput(983.59,660.09)(0.482,-0.762){9}{\rule{0.116pt}{0.700pt}}
\multiput(982.17,661.55)(6.000,-7.547){2}{\rule{0.400pt}{0.350pt}}
\multiput(989.59,650.60)(0.477,-0.933){7}{\rule{0.115pt}{0.820pt}}
\multiput(988.17,652.30)(5.000,-7.298){2}{\rule{0.400pt}{0.410pt}}
\multiput(994.59,641.93)(0.477,-0.821){7}{\rule{0.115pt}{0.740pt}}
\multiput(993.17,643.46)(5.000,-6.464){2}{\rule{0.400pt}{0.370pt}}
\multiput(999.59,633.60)(0.477,-0.933){7}{\rule{0.115pt}{0.820pt}}
\multiput(998.17,635.30)(5.000,-7.298){2}{\rule{0.400pt}{0.410pt}}
\multiput(1004.59,624.60)(0.477,-0.933){7}{\rule{0.115pt}{0.820pt}}
\multiput(1003.17,626.30)(5.000,-7.298){2}{\rule{0.400pt}{0.410pt}}
\multiput(1009.60,614.85)(0.468,-1.212){5}{\rule{0.113pt}{1.000pt}}
\multiput(1008.17,616.92)(4.000,-6.924){2}{\rule{0.400pt}{0.500pt}}
\multiput(1013.59,606.93)(0.477,-0.821){7}{\rule{0.115pt}{0.740pt}}
\multiput(1012.17,608.46)(5.000,-6.464){2}{\rule{0.400pt}{0.370pt}}
\multiput(1018.59,598.60)(0.477,-0.933){7}{\rule{0.115pt}{0.820pt}}
\multiput(1017.17,600.30)(5.000,-7.298){2}{\rule{0.400pt}{0.410pt}}
\multiput(1023.60,588.85)(0.468,-1.212){5}{\rule{0.113pt}{1.000pt}}
\multiput(1022.17,590.92)(4.000,-6.924){2}{\rule{0.400pt}{0.500pt}}
\multiput(1027.60,580.26)(0.468,-1.066){5}{\rule{0.113pt}{0.900pt}}
\multiput(1026.17,582.13)(4.000,-6.132){2}{\rule{0.400pt}{0.450pt}}
\multiput(1031.59,572.60)(0.477,-0.933){7}{\rule{0.115pt}{0.820pt}}
\multiput(1030.17,574.30)(5.000,-7.298){2}{\rule{0.400pt}{0.410pt}}
\multiput(1036.60,562.85)(0.468,-1.212){5}{\rule{0.113pt}{1.000pt}}
\multiput(1035.17,564.92)(4.000,-6.924){2}{\rule{0.400pt}{0.500pt}}
\multiput(1040.60,554.26)(0.468,-1.066){5}{\rule{0.113pt}{0.900pt}}
\multiput(1039.17,556.13)(4.000,-6.132){2}{\rule{0.400pt}{0.450pt}}
\multiput(1044.60,545.85)(0.468,-1.212){5}{\rule{0.113pt}{1.000pt}}
\multiput(1043.17,547.92)(4.000,-6.924){2}{\rule{0.400pt}{0.500pt}}
\multiput(1048.60,536.85)(0.468,-1.212){5}{\rule{0.113pt}{1.000pt}}
\multiput(1047.17,538.92)(4.000,-6.924){2}{\rule{0.400pt}{0.500pt}}
\multiput(1052.60,528.26)(0.468,-1.066){5}{\rule{0.113pt}{0.900pt}}
\multiput(1051.17,530.13)(4.000,-6.132){2}{\rule{0.400pt}{0.450pt}}
\multiput(1056.60,519.85)(0.468,-1.212){5}{\rule{0.113pt}{1.000pt}}
\multiput(1055.17,521.92)(4.000,-6.924){2}{\rule{0.400pt}{0.500pt}}
\multiput(1060.60,510.85)(0.468,-1.212){5}{\rule{0.113pt}{1.000pt}}
\multiput(1059.17,512.92)(4.000,-6.924){2}{\rule{0.400pt}{0.500pt}}
\multiput(1064.61,501.16)(0.447,-1.579){3}{\rule{0.108pt}{1.167pt}}
\multiput(1063.17,503.58)(3.000,-5.579){2}{\rule{0.400pt}{0.583pt}}
\multiput(1067.60,493.85)(0.468,-1.212){5}{\rule{0.113pt}{1.000pt}}
\multiput(1066.17,495.92)(4.000,-6.924){2}{\rule{0.400pt}{0.500pt}}
\multiput(1071.60,484.85)(0.468,-1.212){5}{\rule{0.113pt}{1.000pt}}
\multiput(1070.17,486.92)(4.000,-6.924){2}{\rule{0.400pt}{0.500pt}}
\multiput(1075.61,475.16)(0.447,-1.579){3}{\rule{0.108pt}{1.167pt}}
\multiput(1074.17,477.58)(3.000,-5.579){2}{\rule{0.400pt}{0.583pt}}
\multiput(1078.60,467.85)(0.468,-1.212){5}{\rule{0.113pt}{1.000pt}}
\multiput(1077.17,469.92)(4.000,-6.924){2}{\rule{0.400pt}{0.500pt}}
\multiput(1082.61,458.16)(0.447,-1.579){3}{\rule{0.108pt}{1.167pt}}
\multiput(1081.17,460.58)(3.000,-5.579){2}{\rule{0.400pt}{0.583pt}}
\multiput(1085.60,450.85)(0.468,-1.212){5}{\rule{0.113pt}{1.000pt}}
\multiput(1084.17,452.92)(4.000,-6.924){2}{\rule{0.400pt}{0.500pt}}
\multiput(1089.61,441.16)(0.447,-1.579){3}{\rule{0.108pt}{1.167pt}}
\multiput(1088.17,443.58)(3.000,-5.579){2}{\rule{0.400pt}{0.583pt}}
\multiput(1092.61,432.60)(0.447,-1.802){3}{\rule{0.108pt}{1.300pt}}
\multiput(1091.17,435.30)(3.000,-6.302){2}{\rule{0.400pt}{0.650pt}}
\multiput(1095.60,425.26)(0.468,-1.066){5}{\rule{0.113pt}{0.900pt}}
\multiput(1094.17,427.13)(4.000,-6.132){2}{\rule{0.400pt}{0.450pt}}
\multiput(1099.61,415.60)(0.447,-1.802){3}{\rule{0.108pt}{1.300pt}}
\multiput(1098.17,418.30)(3.000,-6.302){2}{\rule{0.400pt}{0.650pt}}
\multiput(1102.61,407.16)(0.447,-1.579){3}{\rule{0.108pt}{1.167pt}}
\multiput(1101.17,409.58)(3.000,-5.579){2}{\rule{0.400pt}{0.583pt}}
\multiput(1105.61,398.60)(0.447,-1.802){3}{\rule{0.108pt}{1.300pt}}
\multiput(1104.17,401.30)(3.000,-6.302){2}{\rule{0.400pt}{0.650pt}}
\multiput(1108.61,390.16)(0.447,-1.579){3}{\rule{0.108pt}{1.167pt}}
\multiput(1107.17,392.58)(3.000,-5.579){2}{\rule{0.400pt}{0.583pt}}
\multiput(1111.61,381.60)(0.447,-1.802){3}{\rule{0.108pt}{1.300pt}}
\multiput(1110.17,384.30)(3.000,-6.302){2}{\rule{0.400pt}{0.650pt}}
\multiput(1114.61,373.16)(0.447,-1.579){3}{\rule{0.108pt}{1.167pt}}
\multiput(1113.17,375.58)(3.000,-5.579){2}{\rule{0.400pt}{0.583pt}}
\multiput(1117.61,364.60)(0.447,-1.802){3}{\rule{0.108pt}{1.300pt}}
\multiput(1116.17,367.30)(3.000,-6.302){2}{\rule{0.400pt}{0.650pt}}
\multiput(1120.61,356.16)(0.447,-1.579){3}{\rule{0.108pt}{1.167pt}}
\multiput(1119.17,358.58)(3.000,-5.579){2}{\rule{0.400pt}{0.583pt}}
\multiput(1123.61,347.60)(0.447,-1.802){3}{\rule{0.108pt}{1.300pt}}
\multiput(1122.17,350.30)(3.000,-6.302){2}{\rule{0.400pt}{0.650pt}}
\multiput(1126.61,339.16)(0.447,-1.579){3}{\rule{0.108pt}{1.167pt}}
\multiput(1125.17,341.58)(3.000,-5.579){2}{\rule{0.400pt}{0.583pt}}
\put(1129.17,327){\rule{0.400pt}{1.900pt}}
\multiput(1128.17,332.06)(2.000,-5.056){2}{\rule{0.400pt}{0.950pt}}
\multiput(1131.61,322.16)(0.447,-1.579){3}{\rule{0.108pt}{1.167pt}}
\multiput(1130.17,324.58)(3.000,-5.579){2}{\rule{0.400pt}{0.583pt}}
\multiput(1134.61,313.60)(0.447,-1.802){3}{\rule{0.108pt}{1.300pt}}
\multiput(1133.17,316.30)(3.000,-6.302){2}{\rule{0.400pt}{0.650pt}}
\multiput(1137.61,305.16)(0.447,-1.579){3}{\rule{0.108pt}{1.167pt}}
\multiput(1136.17,307.58)(3.000,-5.579){2}{\rule{0.400pt}{0.583pt}}
\put(1140.17,294){\rule{0.400pt}{1.700pt}}
\multiput(1139.17,298.47)(2.000,-4.472){2}{\rule{0.400pt}{0.850pt}}
\multiput(1142.61,288.60)(0.447,-1.802){3}{\rule{0.108pt}{1.300pt}}
\multiput(1141.17,291.30)(3.000,-6.302){2}{\rule{0.400pt}{0.650pt}}
\multiput(1145.61,280.16)(0.447,-1.579){3}{\rule{0.108pt}{1.167pt}}
\multiput(1144.17,282.58)(3.000,-5.579){2}{\rule{0.400pt}{0.583pt}}
\put(1148.17,269){\rule{0.400pt}{1.700pt}}
\multiput(1147.17,273.47)(2.000,-4.472){2}{\rule{0.400pt}{0.850pt}}
\multiput(1150.61,263.60)(0.447,-1.802){3}{\rule{0.108pt}{1.300pt}}
\multiput(1149.17,266.30)(3.000,-6.302){2}{\rule{0.400pt}{0.650pt}}
\put(1153.17,252){\rule{0.400pt}{1.700pt}}
\multiput(1152.17,256.47)(2.000,-4.472){2}{\rule{0.400pt}{0.850pt}}
\multiput(1155.61,246.60)(0.447,-1.802){3}{\rule{0.108pt}{1.300pt}}
\multiput(1154.17,249.30)(3.000,-6.302){2}{\rule{0.400pt}{0.650pt}}
\put(1158.17,235){\rule{0.400pt}{1.700pt}}
\multiput(1157.17,239.47)(2.000,-4.472){2}{\rule{0.400pt}{0.850pt}}
\multiput(1160.61,230.16)(0.447,-1.579){3}{\rule{0.108pt}{1.167pt}}
\multiput(1159.17,232.58)(3.000,-5.579){2}{\rule{0.400pt}{0.583pt}}
\put(1163.17,219){\rule{0.400pt}{1.700pt}}
\multiput(1162.17,223.47)(2.000,-4.472){2}{\rule{0.400pt}{0.850pt}}
\put(1165.17,210){\rule{0.400pt}{1.900pt}}
\multiput(1164.17,215.06)(2.000,-5.056){2}{\rule{0.400pt}{0.950pt}}
\multiput(1167.61,205.16)(0.447,-1.579){3}{\rule{0.108pt}{1.167pt}}
\multiput(1166.17,207.58)(3.000,-5.579){2}{\rule{0.400pt}{0.583pt}}
\put(1170.17,194){\rule{0.400pt}{1.700pt}}
\multiput(1169.17,198.47)(2.000,-4.472){2}{\rule{0.400pt}{0.850pt}}
\put(1172.17,185){\rule{0.400pt}{1.900pt}}
\multiput(1171.17,190.06)(2.000,-5.056){2}{\rule{0.400pt}{0.950pt}}
\multiput(1174.61,180.16)(0.447,-1.579){3}{\rule{0.108pt}{1.167pt}}
\multiput(1173.17,182.58)(3.000,-5.579){2}{\rule{0.400pt}{0.583pt}}
\put(140,1040){\usebox{\plotpoint}}
\put(660,1038.67){\rule{5.059pt}{0.400pt}}
\multiput(660.00,1039.17)(10.500,-1.000){2}{\rule{2.529pt}{0.400pt}}
\put(140.0,1040.0){\rule[-0.200pt]{125.268pt}{0.400pt}}
\put(792,1037.67){\rule{2.891pt}{0.400pt}}
\multiput(792.00,1038.17)(6.000,-1.000){2}{\rule{1.445pt}{0.400pt}}
\put(681.0,1039.0){\rule[-0.200pt]{26.740pt}{0.400pt}}
\put(847,1036.67){\rule{2.409pt}{0.400pt}}
\multiput(847.00,1037.17)(5.000,-1.000){2}{\rule{1.204pt}{0.400pt}}
\put(804.0,1038.0){\rule[-0.200pt]{10.359pt}{0.400pt}}
\put(884,1035.67){\rule{1.927pt}{0.400pt}}
\multiput(884.00,1036.17)(4.000,-1.000){2}{\rule{0.964pt}{0.400pt}}
\put(857.0,1037.0){\rule[-0.200pt]{6.504pt}{0.400pt}}
\put(915,1034.67){\rule{1.686pt}{0.400pt}}
\multiput(915.00,1035.17)(3.500,-1.000){2}{\rule{0.843pt}{0.400pt}}
\put(892.0,1036.0){\rule[-0.200pt]{5.541pt}{0.400pt}}
\put(936,1033.67){\rule{1.445pt}{0.400pt}}
\multiput(936.00,1034.17)(3.000,-1.000){2}{\rule{0.723pt}{0.400pt}}
\put(922.0,1035.0){\rule[-0.200pt]{3.373pt}{0.400pt}}
\put(955,1032.67){\rule{1.445pt}{0.400pt}}
\multiput(955.00,1033.17)(3.000,-1.000){2}{\rule{0.723pt}{0.400pt}}
\put(942.0,1034.0){\rule[-0.200pt]{3.132pt}{0.400pt}}
\put(972,1031.67){\rule{1.445pt}{0.400pt}}
\multiput(972.00,1032.17)(3.000,-1.000){2}{\rule{0.723pt}{0.400pt}}
\put(961.0,1033.0){\rule[-0.200pt]{2.650pt}{0.400pt}}
\put(989,1030.67){\rule{1.204pt}{0.400pt}}
\multiput(989.00,1031.17)(2.500,-1.000){2}{\rule{0.602pt}{0.400pt}}
\put(978.0,1032.0){\rule[-0.200pt]{2.650pt}{0.400pt}}
\put(999,1029.67){\rule{1.204pt}{0.400pt}}
\multiput(999.00,1030.17)(2.500,-1.000){2}{\rule{0.602pt}{0.400pt}}
\put(994.0,1031.0){\rule[-0.200pt]{1.204pt}{0.400pt}}
\put(1009,1028.67){\rule{0.964pt}{0.400pt}}
\multiput(1009.00,1029.17)(2.000,-1.000){2}{\rule{0.482pt}{0.400pt}}
\put(1004.0,1030.0){\rule[-0.200pt]{1.204pt}{0.400pt}}
\put(1023,1027.67){\rule{0.964pt}{0.400pt}}
\multiput(1023.00,1028.17)(2.000,-1.000){2}{\rule{0.482pt}{0.400pt}}
\put(1013.0,1029.0){\rule[-0.200pt]{2.409pt}{0.400pt}}
\put(1031,1026.67){\rule{1.204pt}{0.400pt}}
\multiput(1031.00,1027.17)(2.500,-1.000){2}{\rule{0.602pt}{0.400pt}}
\put(1027.0,1028.0){\rule[-0.200pt]{0.964pt}{0.400pt}}
\put(1040,1025.67){\rule{0.964pt}{0.400pt}}
\multiput(1040.00,1026.17)(2.000,-1.000){2}{\rule{0.482pt}{0.400pt}}
\put(1036.0,1027.0){\rule[-0.200pt]{0.964pt}{0.400pt}}
\put(1048,1024.67){\rule{0.964pt}{0.400pt}}
\multiput(1048.00,1025.17)(2.000,-1.000){2}{\rule{0.482pt}{0.400pt}}
\put(1044.0,1026.0){\rule[-0.200pt]{0.964pt}{0.400pt}}
\put(1056,1023.67){\rule{0.964pt}{0.400pt}}
\multiput(1056.00,1024.17)(2.000,-1.000){2}{\rule{0.482pt}{0.400pt}}
\put(1060,1022.67){\rule{0.964pt}{0.400pt}}
\multiput(1060.00,1023.17)(2.000,-1.000){2}{\rule{0.482pt}{0.400pt}}
\put(1052.0,1025.0){\rule[-0.200pt]{0.964pt}{0.400pt}}
\put(1067,1021.67){\rule{0.964pt}{0.400pt}}
\multiput(1067.00,1022.17)(2.000,-1.000){2}{\rule{0.482pt}{0.400pt}}
\put(1064.0,1023.0){\rule[-0.200pt]{0.723pt}{0.400pt}}
\put(1075,1020.67){\rule{0.723pt}{0.400pt}}
\multiput(1075.00,1021.17)(1.500,-1.000){2}{\rule{0.361pt}{0.400pt}}
\put(1071.0,1022.0){\rule[-0.200pt]{0.964pt}{0.400pt}}
\put(1082,1019.67){\rule{0.723pt}{0.400pt}}
\multiput(1082.00,1020.17)(1.500,-1.000){2}{\rule{0.361pt}{0.400pt}}
\put(1085,1018.67){\rule{0.964pt}{0.400pt}}
\multiput(1085.00,1019.17)(2.000,-1.000){2}{\rule{0.482pt}{0.400pt}}
\put(1078.0,1021.0){\rule[-0.200pt]{0.964pt}{0.400pt}}
\put(1092,1017.67){\rule{0.723pt}{0.400pt}}
\multiput(1092.00,1018.17)(1.500,-1.000){2}{\rule{0.361pt}{0.400pt}}
\put(1089.0,1019.0){\rule[-0.200pt]{0.723pt}{0.400pt}}
\put(1099,1016.67){\rule{0.723pt}{0.400pt}}
\multiput(1099.00,1017.17)(1.500,-1.000){2}{\rule{0.361pt}{0.400pt}}
\put(1102,1015.67){\rule{0.723pt}{0.400pt}}
\multiput(1102.00,1016.17)(1.500,-1.000){2}{\rule{0.361pt}{0.400pt}}
\put(1095.0,1018.0){\rule[-0.200pt]{0.964pt}{0.400pt}}
\put(1108,1014.67){\rule{0.723pt}{0.400pt}}
\multiput(1108.00,1015.17)(1.500,-1.000){2}{\rule{0.361pt}{0.400pt}}
\put(1111,1013.67){\rule{0.723pt}{0.400pt}}
\multiput(1111.00,1014.17)(1.500,-1.000){2}{\rule{0.361pt}{0.400pt}}
\put(1105.0,1016.0){\rule[-0.200pt]{0.723pt}{0.400pt}}
\put(1117,1012.67){\rule{0.723pt}{0.400pt}}
\multiput(1117.00,1013.17)(1.500,-1.000){2}{\rule{0.361pt}{0.400pt}}
\put(1120,1011.67){\rule{0.723pt}{0.400pt}}
\multiput(1120.00,1012.17)(1.500,-1.000){2}{\rule{0.361pt}{0.400pt}}
\put(1123,1010.67){\rule{0.723pt}{0.400pt}}
\multiput(1123.00,1011.17)(1.500,-1.000){2}{\rule{0.361pt}{0.400pt}}
\put(1114.0,1014.0){\rule[-0.200pt]{0.723pt}{0.400pt}}
\put(1129,1009.67){\rule{0.482pt}{0.400pt}}
\multiput(1129.00,1010.17)(1.000,-1.000){2}{\rule{0.241pt}{0.400pt}}
\put(1131,1008.67){\rule{0.723pt}{0.400pt}}
\multiput(1131.00,1009.17)(1.500,-1.000){2}{\rule{0.361pt}{0.400pt}}
\put(1134,1007.67){\rule{0.723pt}{0.400pt}}
\multiput(1134.00,1008.17)(1.500,-1.000){2}{\rule{0.361pt}{0.400pt}}
\put(1126.0,1011.0){\rule[-0.200pt]{0.723pt}{0.400pt}}
\put(1140,1006.67){\rule{0.482pt}{0.400pt}}
\multiput(1140.00,1007.17)(1.000,-1.000){2}{\rule{0.241pt}{0.400pt}}
\put(1142,1005.67){\rule{0.723pt}{0.400pt}}
\multiput(1142.00,1006.17)(1.500,-1.000){2}{\rule{0.361pt}{0.400pt}}
\put(1145,1004.67){\rule{0.723pt}{0.400pt}}
\multiput(1145.00,1005.17)(1.500,-1.000){2}{\rule{0.361pt}{0.400pt}}
\put(1148,1003.67){\rule{0.482pt}{0.400pt}}
\multiput(1148.00,1004.17)(1.000,-1.000){2}{\rule{0.241pt}{0.400pt}}
\put(1137.0,1008.0){\rule[-0.200pt]{0.723pt}{0.400pt}}
\put(1153,1002.67){\rule{0.482pt}{0.400pt}}
\multiput(1153.00,1003.17)(1.000,-1.000){2}{\rule{0.241pt}{0.400pt}}
\put(1155,1001.67){\rule{0.723pt}{0.400pt}}
\multiput(1155.00,1002.17)(1.500,-1.000){2}{\rule{0.361pt}{0.400pt}}
\put(1158,1000.67){\rule{0.482pt}{0.400pt}}
\multiput(1158.00,1001.17)(1.000,-1.000){2}{\rule{0.241pt}{0.400pt}}
\put(1160,999.67){\rule{0.723pt}{0.400pt}}
\multiput(1160.00,1000.17)(1.500,-1.000){2}{\rule{0.361pt}{0.400pt}}
\put(1163,998.67){\rule{0.482pt}{0.400pt}}
\multiput(1163.00,999.17)(1.000,-1.000){2}{\rule{0.241pt}{0.400pt}}
\put(1150.0,1004.0){\rule[-0.200pt]{0.723pt}{0.400pt}}
\put(1167,997.67){\rule{0.723pt}{0.400pt}}
\multiput(1167.00,998.17)(1.500,-1.000){2}{\rule{0.361pt}{0.400pt}}
\put(1170,996.67){\rule{0.482pt}{0.400pt}}
\multiput(1170.00,997.17)(1.000,-1.000){2}{\rule{0.241pt}{0.400pt}}
\put(1172,995.67){\rule{0.482pt}{0.400pt}}
\multiput(1172.00,996.17)(1.000,-1.000){2}{\rule{0.241pt}{0.400pt}}
\put(1174,994.67){\rule{0.723pt}{0.400pt}}
\multiput(1174.00,995.17)(1.500,-1.000){2}{\rule{0.361pt}{0.400pt}}
\put(1165.0,999.0){\rule[-0.200pt]{0.482pt}{0.400pt}}
\end{picture}
\vspace{0.5cm}
\caption{Illustrative plots for the correlation between
${\cal J}$, $\sin^2 2\theta_{\rm sun}$ or $\sin^2 2 \theta_{\rm atm}$ 
and $|V_{e3}|^2$ in the new neutrino mixing scenario.}
\end{figure}
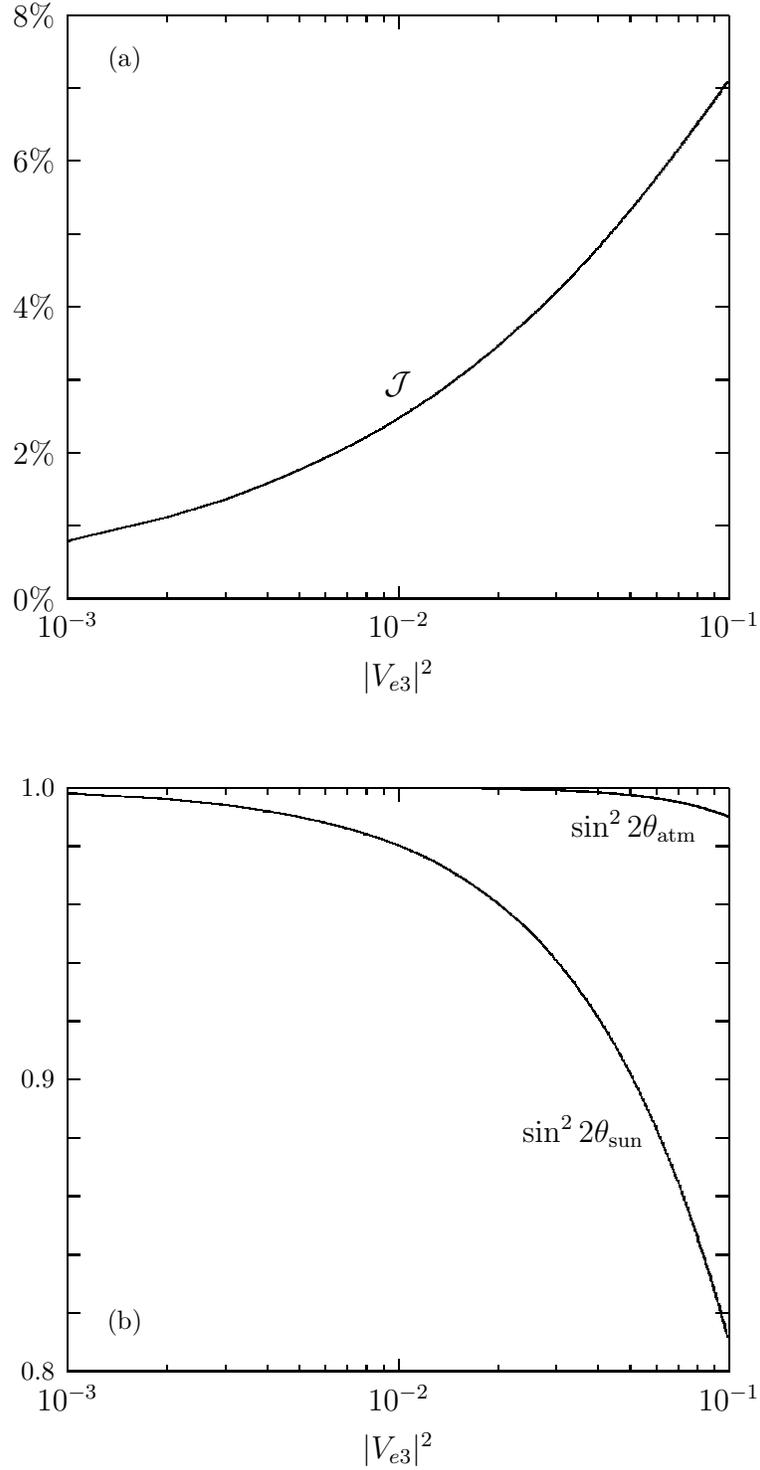

\end{document}